\documentclass[11pt]{article} 
\usepackage{graphicx,floatflt,amssymb,epsf,psfig,rotate} 
\usepackage{color}

\def\ltap{\raisebox{-.4ex}{\rlap{$\sim$}} \raisebox{.4ex}{$<$}} 
\def\gtap{\raisebox{-.4ex}{\rlap{$\sim$}} \raisebox{.4ex}{$>$}} 

\begin{document} 
\begin{flushright} 
SINP/TNP/07-14 \\
HRI-P-07-05-002 \\
CU-Physics-11/2007
\end{flushright} 
 
\vskip 30pt 
 
\begin{center} 
{\Large \bf Extra-dimensional relaxation of the upper limit 
of the \\lightest supersymmetric neutral Higgs mass} \\
\vspace*{1cm} 
\renewcommand{\thefootnote}{\fnsymbol{footnote}} 
{ {\sf Gautam Bhattacharyya${}^1$}, {\sf Swarup Kumar Majee${}^{2,3}$}, 
{\sf Amitava Raychaudhuri${}^{2,3}$} 
} \\ 
\vspace{10pt} 
{\small ${}^{1)}$ {\em Saha Institute of Nuclear Physics, 
1/AF Bidhan Nagar, Kolkata 700064, India} \\
   ${}^{2)}$ {\em Harish-Chandra Research Institute,
Chhatnag Road, Jhunsi, Allahabad  211019, India} \\  
   ${}^{3)}$ {\em Department of Physics, University of Calcutta, 
92 A.P.C. Road, Kolkata 700009, India}}  
 
\normalsize 
\end{center} 

\begin{abstract} 
The upper limit on the mass of the lightest CP-even neutral Higgs in
the minimal supersymmetric standard model is around 135 GeV for soft
supersymmetry breaking masses in the 1 TeV range.  We demonstrate that this
upper limit may be sizably relaxed if supersymmetry is embedded in
extra dimensions.  We calculate, using the effective potential
technique, the radiative corrections to the lightest Higgs mass
induced by the Kaluza-Klein towers of quarks and squarks with one and
two compactified directions. We observe that the lightest Higgs may
comfortably weigh around 200 GeV (300 GeV) with one (two) extra
dimension(s).
 
\vskip 5pt \noindent 
\texttt{PACS Nos:~ 12.60.Jv, 11.10.Kk, 14.80.Cp } \\ 
\texttt{Key Words:~~Supersymmetry, Extra Dimension, Higgs mass}
\end{abstract}

\renewcommand{\thesection}{\Roman{section}} 
\setcounter{footnote}{0} 
\renewcommand{\thefootnote}{\arabic{footnote}} 
 
\section{Introduction} 
The most general symmetries of local relativistic quantum field
theories include supersymmetry, a phenomenological version
\cite{books} of which is awaiting a final judgement within the next
few years as the Large Hadron Collider (LHC) turns on. Indeed, one of
the most coveted targets of the LHC is to capture the Higgs boson, and
supersymmetry, admitting chiral fermions together with their scalar
partners in the same representations, tacitly provides a rationale for
treating the Higgs as an elementary object
\cite{Ellis:2007wa}. Furthermore, through the removal of the
quadratic divergence that plagues the ordinary Higgs mass,
phenomenological supersymmetry has emerged as a leading candidate of
physics beyond the standard model (SM).  A key signature of the
minimal version of supersymmetry is that the lightest Higgs boson mass
obeys an upper bound ($\sim$ 135 GeV, see below) -- a prediction which
will be put to test during the LHC run. Now, supersymmetry is an
integral part of string theory which attempts to provide a quantum
picture of all interactions.  Since string theory is intrinsically a
higher dimensional theory, a reanalysis of some 4-dimensional (4d)
supersymmetric wisdom in the backdrop of extra dimensions might
provide important clues to our search strategies.  Considering that
the Higgs is the most-wanted entity at the LHC, in this paper we
address the following question which we believe is extremely timely:
{\em What is the upper limit of the lightest CP-even neutral Higgs
mass if the minimal supersymmetric standard model (MSSM) is embedded
in extra dimensions}?  We consider the embedding first in one and then
in two extra dimensions.

Let us first discuss why this is an important issue. Recall that MSSM
has two Higgs doublet superfields ($H_1$ and $H_2$), and supersymmetry
does not allow the scalar potential to have independent quartic
couplings. Gauge interactions generate them through supersymmetry
breaking $D$-terms and the effective quartic interactions are written
in terms of the gauge couplings.  This makes the Higgs spectrum
partially predictive, in the sense that at the tree level the lightest
neutral Higgs ($h$) weighs less than $M_Z$ ($m_h^2 < M_Z^2
\cos^2 2\beta$, where $\tan \beta$ is the ratio of two vacuum expectation
values (VEVs)). However, $m_h$ receives quantum corrections which, due
to the large top quark Yukawa coupling and for heavy stop squarks, can
become as large as $\Delta m_h^2 \sim (3G_F m_t^4/\sqrt{2}\pi^2) \ln
(m_{\tilde{t}}^2/m_t^2)$, where $m_{\tilde{t}}$ is an average stop
squark mass \cite{radcorr1,radcorr2}.  The upper limit on $m_h$ is
then pushed to around 135 GeV for squark mass in the ${\cal{O}}$(TeV)
range. Notice that the non-observation of the Higgs boson at LEP2 has
already set a lower limit $m_h > 114.5$ GeV
\cite{Barate:2003sz,Schael:2006cr}, which is satisfied only if a
sizable quantum correction elevates the Higgs mass beyond the tree
level upper limit of $M_Z$. This implies (i) lower values of $\tan
\beta$, which is usually chosen in the range $1 < \tan\beta <
m_t/m_b$, are disfavoured, and (ii) the squark mass $m_{\tilde{t}}$
has to be in the TeV range, which also sets the scale of a generic
soft supersymmetry breaking mass $M_S$.  The MSSM prediction of a
light Higgs is also in line with the indication coming from
electroweak precision tests that the neutral Higgs should weigh below
199 GeV\footnote{This indirect upper limit as well as the LEP2 direct
search lower limit of $m_h >$ 114.5 GeV apply, strictly speaking, for
the SM Higgs.  However, in the `decoupling limit' of the MSSM (large
$m_A$ leading to full-strength $ZZh$ coupling), which is the region of
interest in the present paper, the above limits continue to hold.}
\cite{Alcaraz:2006mx}.  The so called `little hierarchy' problem then
arises out of an order of magnitude mass splitting between the Higgs
and the superparticles.

Adding a gauge singlet superfield ($N$) in the MSSM spectrum and
coupling it with $H_{1,2}$ via the superpotential $\lambda N H_1 H_2$
helps to ease the tension. Not only does this next to minimal version
of supersymmetry (the so called NMSSM \cite{Ellwanger:1999ji}) help to
address the `$\mu$ problem', it also generates a tree level quartic
coupling in the scalar potential which modifies the tree level upper
limit on $m_h$ through $m_h^2 < M_Z^2 \cos^2 2\beta [1 + 2\lambda^2
\tan^2 2\beta/(g^2+g'^2)]$ (see
\cite{Drees:1988fc}). Assuming $\lambda$ to be in the perturbative regime,
i.e., $\lambda \sim g,g'$, one basically obtains a new contribution
$\sim M_Z^2 \sin^2 2\beta$ to the tree level $m_h^2$. This way the low
$\tan\beta$ regime can be revived. Since many supersymmetric couplings
depend on $\tan\beta$, search strategies alter in a significant way if
the disfavoured low $\tan\beta$ region is thus
resurrected\footnote{Low $\tan\beta$ is preferred by electroweak
baryogenesis as well
\cite{ewb}.}${}^,$\footnote{The constraint arising from perturbativity of
couplings can be evaded if the Higgs is charged under an
asymptotically free gauge group \cite{Batra:2003nj}.}.

In this paper we adopt a different approach which also revives the low
$\tan\beta$ region. We stick to the MSSM particle content, but embed
it in a higher dimension compactified at the inverse TeV scale
\cite{dienes}. Although we argued in the beginning that string theory
provides a rationale for linking the two ideas, namely, supersymmetry
and extra dimension, establishing any rigourous connection between the
two at the level of phenomenological models is still a long shot. Here
we take a `bottom-up' approach: we first outline what has already been
studied in the phenomenological context of TeV scale extra-dimensional
scenarios, and then illustrate what we aim to achieve in this paper.
\begin{enumerate}
\item 
Consider first scenarios with one extra dimension (with inverse radius
of compactification around a TeV) but without supersymmetry.  A
typical model is the universal extra-dimensional scenario (UED) where
all particles access the extra dimension \cite{acd}.  Constraints on
this scenario from $g-2$ of the muon \cite{nath}, flavour changing
neutral currents \cite{chk,buras,desh}, $Z \to b\bar{b}$ decay
\cite{santa}, the $\rho$ parameter \cite{acd,appel-yee}, other
electroweak precision tests \cite{ewued}, implications from hadron
collider studies \cite{collued}, etc. imply that $R^{-1}~\gtap~300$
GeV.  A recent inclusive $\bar{B} \to X_s \gamma$ analysis sets a
stronger constraint $R^{-1}~\gtap~600$ GeV \cite{Haisch:2007vb}.
These scenarios are motivated from many phenomenological angles.  They
could lead to a new mechanism of supersymmetry breaking
\cite{antoniadis1}, address the fermion mass hierarchy in an
alternative way \cite{Arkani-Hamed:1999dc}, provide a cosmologically
viable dark matter candidate \cite{Servant:2002aq}, stimulate power
law renormalization group running \cite{dienes,blitzkrieg}, admit
substantial evolution of neutrino mixing angles defined through an
effective Majorana neutrino mass operator \cite{Deandrea:2006mh},
etc\footnote{Ultraviolet cutoff sensitivity in different kinds of TeV
scale extra-dimensional models has been dealt in \cite{db}.}.

\item 
Our object of interest is a supersymmetric theory (e.g. MSSM) but
embedded in a higher dimension. Here we ask the following question:
{\em What would be the shift in the Higgs mass due to radiative
effects induced by extra dimensions}?  In the kind of scenarios we
consider, the SM bosons along with their superpartners access the
higher dimensional bulk. Additionally, SM fermions of one or more
generations together with their superpartners also do so. From a 4d
perspective, all the states which access the bulk will have
Kaluza-Klein (KK) towers. The zero modes, i.e., those states which do
not have any momenta along the extra coordinates, are identified with
the standard 4d MSSM spectra. Now, not only the top quark and the stop
squarks would contribute to the radiative correction to $m_h^2$, their
KK partners would do so as well. As it turns out, the radiative
correction driven by the KK states has the same sign as the one from
the zero modes. As a result, $\Delta m_h^2$ becomes larger and thus
the upper limit on $m_h$ is pushed to higher values beyond the usual
4d MSSM limit of around 135 GeV. As we shall see, in the absence of any
left-right scalar mixing, the new contribution coming from KK modes is
to a good approximation proportional to $R^2 (m_{\tilde{t}}^2 -
m_t^2)/n^2$. This fits our intuition that the KK contribution falls
with higher KK modes and vanishes both when $R \to 0$ and in the limit
of exact supersymmetry. We can interpret the result in two
ways. Either, we take large $\tan\beta$ and ${\cal{O}}$(TeV) squark
mass that yielded the 4d supersymmetry limit $\sim$ 135 GeV, in which case
the new upper limit shoots up by several tens of GeV. Or, we may admit
lower $\tan\beta$ and/or accommodate lighter zero mode squarks which were
hitherto disfavoured in the 4d context.  Either way, the Higgs
phenomenology gets an interesting twist which is intuitively
comprehensible and analytically tractable, owing largely due to the
fact that we are here dealing with only {\em one additional}
parameter, namely, the radius of compactification. Moreover, the top
quark mass which appears with fourth power in the expression of
$\Delta m_h^2$ is now known to a precision better than ever ($m_t =
170.9 \pm 1.8$ GeV
\cite{Brubaker:2006xn}).
\end{enumerate} 

As mentioned before, we have considered the embedding of 4d supersymmetry in
one as well as two extra dimensions. There are quite a few advantages of
considering a 6d gauge theory \cite{sixd} even in a non-supersymmetric
scenario: (i) number of fermion generations is restricted to three, or a
multiple of it, if the global SU(2) gauge anomaly has to cancel
\cite{anomaly-can}, (ii) proton decay is adequately suppressed, which is
difficult to achieve in 5d UED, thanks to a discrete symmetry that
survives as a subgroup of the 6d Lorentz group \cite{pdecay},
(iii) observed neutrino masses and mixings can be nicely
explained \cite{neutrino}, and (iv) KK vector modes offer better
opportunities to be explored \cite{sixdcoll}. We shall see that
qualitatively the KK contributions to the radiative corrections
of $m_h$ from 5d and 6d theories are similar, the quantitative
estimates differ due to the different density of KK states in the
two cases. In 5d, the KK states are spaced as $n/R$ (modulo their
zero mode masses) where $n$, an integer, is the KK number,
whereas in 6d, a similar expression holds except $n^2 \Rightarrow
j^2+k^2$, where $j$ and $k$ are two different sets of KK numbers
corresponding to the two compactified directions.

Section II is basically a review of the standard derivation of
the upper limit of the lightest neutral Higgs in conventional 4d
MSSM in the effective potential approach. This paves the way, in
Section III, to upgrade the above derivation for accommodating
contributions from the KK modes of the top quark and squarks in
5d and 6d scenarios. In Section IV, we shall comment on the
numerical impact of the higher KK modes on the lightest neutral
Higgs mass and its consequences. We shall draw our conclusion in
the final section.

\section{MSSM neutral Higgs spectrum in 4 dimensions}
\subsection{Tree level mass relations}
MSSM requires two Higgs doublets for three good reasons: (i) to avoid massless
charged degrees of freedom, (ii) to maintain analyticity of the
superpotential, and (iii) to keep the theory free from chiral anomaly, which
requires two Higgs doublets with opposite hypercharges.

We denote these two complex scalar doublets as
\begin{eqnarray}
&& {H_1} =\left(\begin{array}{c} {H_1}^0\\ {H_1}^- \end{array}\right),
~~~~~ {H_2}=\left(\begin{array}{c} {H_2}^+\\ {H_2}^0
\end{array}\right), \label{higgs}
\end{eqnarray}
whose SU(2)$\times$ U(1) quantum numbers are (2,$-$1) and (2,$+$1)
respectively. $H_1^0$ couples with down-type quarks and charged leptons, while
$H_2^0$ couples with up-type quarks. This guarantees natural suppression of
flavour-changing neutral currents in the limit of exact supersymmetry.   
The tree level potential involving these two doublets 
is given by 
\begin{equation}
V = m_1^2|H_1|^2 + m_2^2|H_2|^2 + m_{12}^2(H_1H_2 + {\rm h.c}) +
{1 \over 8}g^2(H_2^{\dagger}\sigma^a H_2 +H_1^{\dagger}\sigma^a H_1)^2 
+{1 \over 8}{g^\prime}^2(|H_2|^2-|H_1|^2 )^2 , \label{pot1}
\end{equation}
where $m_1^2$, $m_2^2$ and $m_{12}^2$ are soft supersymmetry breaking mass
parameters, $g$ and $g^{\prime}$ are the SU(2) and U(1) gauge couplings, and
$\sigma^a$ $(a=1,2,3)$ are the Pauli matrices. Note that the quartic coupling
is related to the gauge couplings. The part involving the neutral fields is
given by
\begin{equation}
V_0 = m_1^2|H_1^0|^2 + m_2^2|H_2^0|^2 - m_{12}^2(H_1^0 H_2^0 + {\rm h.c}) +
{1 \over 8}(g^2+{g^\prime}^2)(|H_1^0|^2-|H_2^0|^2)^2 . \label{pot2}
\end{equation}
After spontaneous symmetry breaking the minimum of $V_0$ involves the
following two VEVs: $\langle H_1^0\rangle = v_1$ and $\langle
H_2^0\rangle = v_2$.  The combination $v = \sqrt{v_1^2+v_2^2} =
(\sqrt{2} G_F)^{-1/2} \simeq 246$ GeV sets the Fermi scale.

Now, out of the eight degrees of freedom contained in the two Higgs
doublets three are absorbed as the longitudinal modes of the $W$ and the
$Z$ bosons, while the remaining five modes appear as physical states. Of
these five states, two are charged $(H^\pm)$ and three are neutral ($h, H,
A$). Our present concern is the neutral sector of which ($h,H$) are
CP-even, while $A$ is CP-odd. From the separate diagonalisation of the
CP-odd and CP-even neutral mass matrices two important relations
emerge:
\begin{eqnarray}  
\label{atreemass}
m_A^2 & = & \frac{2 m_{12}^2}{\sin 2\beta} ~,~~~{\rm where}~~ 
\tan\beta = \frac{v_2}{v_1} ,\\
\label{htreemass}
m^2_{h,H} & = & {1\over 2}
\left[m_A^2 + M_Z^2 \mp \sqrt{(m_A^2 + M_Z^2 )^2 - 
4 m_A^2 M_Z^2 \cos^2 2\beta} \right] , 
\end{eqnarray} 
where, by definition, $h$ is the lighter of the two CP-even Higgs. 
These in turn give rise to the following sum rule and inequality: 
\begin{eqnarray}
\label{sumrule}
& & m_h^2 + m_H^2  =  m_A^2 + M_Z^2 \\
\label{inequal} 
& & m_h  \leq  {\rm min}~(m_A, M_Z) |\cos 2\beta| \leq {\rm min}~(m_A, M_Z) , 
\end{eqnarray}
i.e., at the tree level (i) the lighter of the two CP-even Higgs ($h$) weighs
less than $M_Z$, and (ii) the CP-odd Higgs ($A$) is heavier than $h$ but
lighter than $H$.

\subsection{Radiative corrections}
We shall now discuss how the above tree level relations are affected
by quantum loops \cite{radcorr1,radcorr2}. We shall confine our
discussion on the correction to $m_h$ only, and that too at the
one-loop level. We note two important points:
\begin{enumerate}
\item Radiative corrections to $m_h$ are dominated by the top quark Yukawa
coupling ($h_t$) and the masses of the stop squarks ($\tilde{t}_1$,
$\tilde{t}_2$). For large values of $\tan \beta$, the
contributions from the $b$-quark sector also assume significance.
We shall ignore loop contributions mediated by lighter quarks
or the gauge bosons.
\item The tree level Higgs mass is protected by supersymmetry. In the limit of
exact supersymmetry, the entire quantum correction vanishes. So
radiative corrections to $m_h$ will be controlled by $M_S$.
\end{enumerate}

Three different approaches have been adopted in the literature to
calculate the radiative corrections to $m_h$: (i) effective potential
technique, (ii) direct diagrammatic calculations, and (iii)
renormalisation group (RG) method, assuming $M_S\gg M_Z$ and fixing
the quartic coupling proportional to $(g^2+g'^2)$ at that scale and
then evolving down to weak scale. In this paper, we shall follow the
effective potential approach primarily for the sake of conveniently
including the effect of new physics later.

We first start with an RG-improved tree level potential $V_0(Q)$ which
contains running masses $m_i^2 (Q)$ and running gauge couplings
$g_i(Q)$.  The full one-loop effective potential is now given by
\begin{equation}
V_1(Q) = V_0(Q) + \Delta V_1(Q) 
\label{epa1}, 
\end{equation}
where, in terms of the field dependent masses $M(H)$, 
\begin{equation}
\Delta V_1(Q) = {1 \over {64\pi^2}}{\rm Str} M^4(H) 
\left \{\ln{M^2(H)\over {Q^2}} - {3 \over 2} \right \}.
\label{epa2} 
\end{equation}
The $Q$-dependence of $\Delta V_1(Q)$ cancels against that of $V_0(Q)$ making
$V_1(Q)$ independent of $Q$ up to higher loop orders.  The supertrace in
Eq.~(\ref{epa2}), defined through
\begin{equation}
{\rm Str} f(m^2) = \sum_i (-1)^{2J_i} (2J_i+1) f(m_i^2), 
\end{equation}
has to be taken over all members of a supermultiplet
and where $m_i^2 \equiv m_i^2(H)$ is the field-dependent mass
eigenvalue of the particle $i$ with spin $J_i$. 
As an example, the contribution from the chiral multiplet containing the top
quark and squarks is given by 
\begin{equation}
\label{epatop}
\Delta V_t = {3 \over {32\pi^2}}
\left\{ m_{\tilde t_1}^4 \left(\ln{m_{\tilde
    t_1}^2\over{Q^2}}-{3\over2}\right)
      + m_{\tilde t_2}^4 \left(\ln{m_{\tilde
          t_2}^2\over{Q^2}}-{3\over2}\right)
      -2 m_t^4 \left(\ln{m_t^2\over{Q^2}}-{3\over2}\right)\right\}, 
\end{equation}
where the overall factor of 3 comes from colour.  Note that $m_{\tilde{t}_i}$
and $m_t$ in Eq.~(\ref{epatop}) are field dependent masses.  Even though $h_b
\ll h_t$, the contribution from the bottom supermultiplet turns out to be
numerically significant in the large $\tan\beta$ region.  $\Delta V_b$
can be written analogously to $\Delta V_t$ with the appropriate
replacements of top and stop masses by bottom and sbottom masses
respectively.

We now explicitly write down the field dependent mass terms. This
simply means a replacement of $v_i$ by $H_i^0$ ($i=1,2$) wherever
$v_i$ appear in the expression of masses. The field dependent top and
bottom quark masses are given by
\begin{equation} 
\label{mthmbh}
m_t^2(H) = h_t^2 |H_2^0|^2 ~;~ m_b^2(H) = h_b^2 |H_1^0|^2. 
\end{equation}
The field dependent stop and sbottom squark mass matrices are written as 
\begin{eqnarray}
&& {M_{\tilde t}^2} (H) =
\left(\begin{array}{cc} 
m_Q^2 + h_t^2|H^0_2|^2 & h_t(A_t H^0_2+\mu{H_1^0}^*)\\  
h_t(A_t{H^0_2}^* +\mu H_1^0) & m_U^2 + h_t^2|H^0_2|^2   
\end{array}\right), 
\label{tsquark}
\end{eqnarray} 
and 
\begin{eqnarray}
&& {M_{\tilde b}^2} (H) =
\left(\begin{array}{cc} 
m_Q^2 + h_b^2|H^0_1|^2 & h_b(A_b H^0_1+\mu{H_2^0}^*)\\  
h_b(A_b{H^0_1}^* +\mu H_2^0) & m_D^2 + h_b^2|H^0_1|^2   
\end{array}\right).  
\label{bsquark}
\end{eqnarray} 
In Eqs.~(\ref{tsquark}) and (\ref{bsquark}) $m_Q$, $m_U$ and $m_D$ are
soft supersymmetry breaking masses, $A_t$ and $A_b$ are trilinear soft
supersymmetry breaking mass dimensional couplings, and $\mu$ is the
supersymmetry preserving mass dimensional parameter connecting $H_1$
and $H_2$ in the superpotential. We take both trilinear and the $\mu$
couplings to be real. We have neglected the $D$-term contributions which
are small, being proportional to gauge couplings. The squark masses
appearing in Eq.~(\ref{epatop}) are obtained from the diagonalisation
of Eq.~(\ref{tsquark}).

We now consider the radiative correction to the CP-odd scalar mass
matrix. The one-loop corrected mass matrix square, obtained by taking double
derivatives of the full potential with respect to the pseudo-scalar
excitations, can be written as
\begin{eqnarray}
{\cal{M}}^2_{({\rm odd})} = 
\left(\begin{array} {cc} {\rm tan}\beta & 1 \\ 
1 & {\rm cot}\beta \end{array} \right) (m_{12}^2 + \Delta) .
\label{cpodd1}
\end{eqnarray}
The radiative corrections generated as a consequence of
supersymmetry breaking are contained in $\Delta = \Delta^{\rm
t} + \Delta^{\rm b}$, which is given by 
\begin{equation} 
\Delta^{\rm t(b)} = - {3\over{32\pi^2}} {h_{t(b)}^2\mu A_{t(b)}
\over\left[{m^2_{{\tilde t_1}({\tilde b_1})} -
m^2_{{\tilde t_2}({\tilde b_2})}}\right]}
\left[f\left(m^2_{{\tilde t_1}({\tilde b_1})}\right)-
f\left(m^2_{{\tilde t_2}({\tilde b_2})}\right)\right]
\end{equation}
where 
\begin{equation}
f(m^2) = 2m^2\left(\ln{m^2\over{Q^2}} - 1 \right) .
\end{equation}
The zero eigenvalue corresponds to the massless Goldstone boson which is eaten
by the $Z$ boson. The massive state is the pseudo-scalar $A$ whose radiatively
corrected mass square is given by 
\begin{equation}
m_A^2 = \frac{2(m_{12}^2 + \Delta)} {\sin 2\beta} . 
\label{cpodd2}
\end{equation}
The $Q$-dependence of $m_A$ cancels in Eq.~(\ref{cpodd2}) up to
one-loop order. In any case, we shall treat the radiatively corrected
$m_A$ as an input parameter.

Now we are all set to calculate the radiative corrections in the neutral
CP-even mass eigenvalues. The one-loop corrected mass matrix square is  
obtained by taking double derivatives of the full potential
with respect to the scalar excitations and is given by 
\begin{eqnarray}
\label{msqeven}
{\cal{M}}^2_{({\rm even})}  =  
\left(\begin{array} {cc} 
M_Z^2 \cos^2 \beta + m_A^2 \sin^2 \beta & 
-(m_A^2 + M_Z^2) \sin\beta \cos\beta \\ 
-(m_A^2 + M_Z^2) \sin\beta \cos\beta &  
M_Z^2 \sin^2 \beta + m_A^2 \cos^2 \beta
\end{array}\right) 
+ {3\over{4\pi^2 v^2}}
\left(\begin{array}{cc} \Delta_{11} & \Delta_{12} \\ 
\Delta_{12} & \Delta_{22}\end{array}\right) ,
&&  \label{cpeven}
\end{eqnarray}
where $\Delta_{ij} = \Delta^{\rm t}_{ij} + \Delta^{\rm b}_{ij}$. The
individual $\Delta_{ij}$'s are explicitly written below: 
\begin{eqnarray}
\label{delta}
\Delta_{11}^t &=& {m_t^4\over{{\sin}^2\beta}}\left(\mu (A_t+\mu {\rm
  cot}\beta)\over{m_{\tilde t_1}^2 - m_{\tilde t_2}^2}\right)^2g(m_{\tilde
  t_1}^2,m_{\tilde t_2}^2) ,\nonumber \\
\Delta_{12}^t &=& {m_t^4\over{{\sin}^2\beta}}{\mu (A_t+\mu {
\cot}\beta)\over{m_{\tilde t_1}^2 - m_{\tilde t_2}^2}}\left[{
\ln}{{m_{\tilde t_1}^2}\over{m_{\tilde t_2}^2}}+{A_t(A_t+\mu {
\cot}\beta)\over{m_{\tilde t_1}^2 - m_{\tilde t_2}^2}} g(m_{\tilde
t_1}^2,m_{\tilde t_2}^2)\right] ,\nonumber\\
\Delta_{22}^t& =& {m_t^4\over{\sin^2\beta}}\left[\ln{{m_{\tilde
t_1}^2}{m_{\tilde t_2}^2}\over{m_t^4}} + {2A_t(A_t+\mu {
\cot}\beta)\over{m_{\tilde t_1}^2 -m_{\tilde t_2}^2}} {\ln}{{m_{\tilde
t_1}^2}\over{m_{\tilde t_2}^2}} + 
\left(A_t(A_t+\mu {\cot}\beta)\over{m_{\tilde t_1}^2 -
m_{\tilde t_2}^2}\right)^2g(m_{\tilde t_1}^2,m_{\tilde t_2}^2)\right] , 
\nonumber\\
\Delta_{11}^b& =& {m_b^4\over{{\cos}^2\beta}}\left[{\ln}{{m_{\tilde
b_1}^2}{m_{\tilde b_2}^2}\over{m_b^4}} + {2A_b(A_b+\mu {
\tan}\beta)\over{m_{\tilde b_1}^2 -m_{\tilde b_2}^2}} {\ln}{{m_{\tilde
b_1}^2}\over{m_{\tilde b_2}^2}} + 
\left(A_b(A_b+\mu \tan\beta)\over{m_{\tilde b_1}^2 -
m_{\tilde b_2}^2}\right)^2g(m_{\tilde b_1}^2,m_{\tilde b_2}^2)\right] ,
\nonumber \\
\Delta_{12}^b &=& {m_b^4\over{{\cos}^2\beta}}{\mu (A_b+\mu {
\tan}\beta)\over{m_{\tilde b_1}^2 - m_{\tilde b_2}^2}}\left[{
\ln}{{m_{\tilde b_1}^2}\over{m_{\tilde b_2}^2}}+{A_b(A_b+\mu {
\tan}\beta)\over{m_{\tilde b_1}^2 - m_{\tilde b_2}^2}} g(m_{\tilde
b_1}^2,m_{\tilde b_2}^2)\right] , \\
\Delta_{22}^b &=& {m_b^4\over{{\cos}^2\beta}}\left(\mu (A_b+\mu {
\tan}\beta)\over{m_{\tilde b_1}^2 - m_{\tilde b_2}^2}\right)^2 g(m_{\tilde
b_1}^2,m_{\tilde b_2}^2) .\nonumber 
\end{eqnarray}
where  
\begin{equation}
g(m_1^2,m_2^2)=2 - {{m_1^2+m_2^2}\over{m_1^2-m_2^2}}\ln{m_1^2\over m_2^2} .
\end{equation}
Two points deserve mention at this stage: 
\begin{enumerate}
\item While the leading log contribution appears in $\Delta_{22}$ for the top
sector, the same appears in $\Delta_{11}$ for the bottom sector. This
happens because the right-handed top quark couples to $H_2$ while the
right-handed bottom quark couples to $H_1$. In the absence of any
left-right scalar mixing, these leading logs are the only radiative
contributions.
\item Ignoring the left-right scalar mixing, the radiative shift to the Higgs
mass square coming from the top-stop sector turns out to be $\Delta
m_h^2 = (3/4\pi^2 v^2) \Delta^{t}_{22} \sin^2\beta \sim (3
m_t^4/2\pi^2 v^2)
\ln (m_{\tilde{t}}^2/m_t^2)$, where $m_{\tilde{t}} = \sqrt{m_{\tilde{t}_1}
m_{\tilde{t}_2}}$ is an average stop mass. This is the expression we
quoted in the Introduction.
\end{enumerate} 

\section {Radiative corrections due to extra dimensions}
Let us first consider just one extra dimension which is compactified
on a circle of radius $R$. We further consider a $Z_2$ orbifolding
identifying $y \to -y$, where $y$ denotes the compactified coordinate.
The orbifolding is crucial for reproducing the chiral zero modes of
the observed fermions. After the compactified direction is integrated
out, the 4d Lagrangian can be written in terms of the zero modes and
their KK partners. For illustration, we first take a
non-supersymmetric scenario and look into the KK mode expansion of
gauge boson, scalar and fermion fields from a 4d perspective. Each
component of a 5d field is either even or odd under $Z_2$. 
The KK expansions are given by,
\begin{eqnarray}
\label{fourier}
A_{\mu}(x,y)&=&\frac{\sqrt{2}}{\sqrt{2\pi
R}}A_{\mu}^{(0)}(x)+\frac{2}{\sqrt{2\pi
R}}\sum^{\infty}_{n=1}A_{\mu}^{(n)}(x)\cos\frac{ny}{R},~~~~
A_5(x,y) = \frac{2}{\sqrt{2\pi
R}}\sum^{\infty}_{n=1}A_5^{(n)}(x)\sin\frac{ny}{R}, \nonumber\\
\phi(x,y)&=&\frac{\sqrt{2}}{\sqrt{2\pi
R}}\phi^{(0)}(x)+\frac{2}{\sqrt{2\pi
R}}\sum^{\infty}_{n=1}\phi^{(n)}(x)\cos\frac{ny}{R}, \nonumber \\
\mathcal{Q}(x,y)&=&\frac{\sqrt{2}}{\sqrt{2\pi
R}}\bigg[{\pmatrix{t\cr b}}_{L}(x)+\sqrt{2}\sum^{\infty}_{n=1}\Big\{
\mathcal{Q}^{(n)}_{L}(x)\cos\frac{ny}{R}+
\mathcal{Q}^{(n)}_{R}(x)\sin\frac{ny}{R}\Big\}\bigg], \\
\mathcal{T}(x,y)&=&\frac{\sqrt{2}}{\sqrt{2\pi
R}}\bigg[t_{R}(x)+\sqrt{2}\sum^{\infty}_{n=1}\Big\{
\mathcal{T}^{(n)}_{R}(x)\cos\frac{ny}{R}+
\mathcal{T}^{(n)}_{L}(x)\sin\frac{ny}{R}\Big\}\bigg], \nonumber\\
\mathcal{B}(x,y)&=&\frac{\sqrt{2}}{\sqrt{2\pi
R}}\bigg[b_{R}(x)+\sqrt{2}\sum^{\infty}_{n=1}\Big\{
\mathcal{B}^{(n)}_{R}(x)\cos\frac{ny}{R}+
\mathcal{B}^{(n)}_{L}(x)\sin\frac{ny}{R}\Big\}\bigg]. \nonumber
\end{eqnarray}
The complex scalar field $\phi (x,y)$ and the gauge boson $A_\mu(x,y)$
are $Z_2$-even fields with their zero modes identified with the SM
scalar doublet and a SM gauge boson respectively. The field $A_5(x,y)$
is a real pseudoscalar field transforming in the adjoint representation of
the gauge group, and does not have any zero mode. The fields
$\mathcal{Q}$, $\mathcal{T}$, and $\mathcal{B}$ denote the 5d quark
SU(2) doublet and SU(2) singlet states of a given generation
(e.g. third generation) whose zero modes are identified with the 4d SM
quark states. We draw attention to two points at this stage: (i) even
though the $Z_2$ orbifolding renders the zero mode fermions to be
chiral, the KK fermions are vector-like; (ii) the KK number $n$ is
conserved at all tree level vertices, while what actually remains
conserved at all order is the KK parity, defined as $(-1)^n$.  As is
well known, the kinetic terms in the extra dimension give rise to the
KK masses in 4d. For a flat extra dimension the KK masses are added in
quadrature with the corresponding zero mode masses both for fermions
and bosons. A generic expression for the $n$-th mode mass,
where $m_0$ is the zero mode mass, is given by
\begin{equation}
\label{kkmass}
m_n^2 = m_0^2 + \frac{n^2}{R^2}, \;\;\;\;\; n = 0,1,2,\ldots
\end{equation}

We now discuss the supersymmetric version of the theory. A 5d $N=1$
supersymmetry from a 4d perspective appears as two $N=1$
supersymmetries forming an $N=2$ theory. For the details of the
hypermultiplet structures of this theory, we refer the readers to
\cite{dienes}. Our concern in this paper is to calculate the
radiative contribution to $m_h$ coming from the KK partners of
particles and superparticles. We now proceed through the
following steps.
\begin{enumerate}  
\item Let us first recall that the $N=2$ supersymmetry prohibits
any bulk Yukawa interaction involving three chiral multiplets. The
Yukawa interaction is considered to be localised at a brane, like $-
(h_{t5}/\Lambda^{3/2})\int d^4x~ \int dy~
\delta(y) \int d^2\theta~ ({\cal{H}}_2{\cal{Q}}{\cal{T}}~+~{\rm
h.c.})$, where the residual supersymmetry is that of $N=1$,
$h_{t5}$ is a dimensionless Yukawa coupling in 5d and $\Lambda$
the cutoff scale. This localisation has a consequence in the
counting of KK degrees of freedom that contribute to the Higgs
mass radiative correction. The delta function ensures that those
fields which accompany the sine function after Fourier
decomposition do not sense the Yukawa interaction.

\item As in the case of 4d (zero mode) supersymmetry, here too
the dominant effect arises solely from the third generation
quark superfields, only that now we have to include the contributions
from their KK towers. We shall continue to ignore contributions 
from the gauge interactions or those from the first two quark
families, as they are not numerically significant. We might as well
formulate a scheme in which the first two generation of matter
superfields are brane-localised and {\em only} the third generation
superfields access the bulk\footnote{If all the three matter
generations are bulk fields, then the theory become non-perturbative
too soon, unless $1/R > 5.0 \times 10^{10}$ GeV \cite{blitzkrieg}. If
only one generation accesses the bulk and the other two are confined
to a brane, then the validity of the theory extends further, allowing
even a perturbative gauge coupling unification, we checked,
around $E \sim 40/R$.}. Keeping this in mind, we displayed the
Fourier decomposition of only the third generation superfields in
Eq.~(\ref{fourier}).

\item In our scheme $M_S$ and $R$ are independent parameters, although
we take them to be of the same order\footnote{This is in contrast to
other higher dimensional supersymmetric scenarios in which both the
superpartner masses and the scale of electroweak symmetry breaking
arising from quantum loops are set by $1/R$, where $R$ is the distance
between the brane at which top quark Yukawa coupling is localised and
the brane where supersymmetry is broken \cite{arkbar}. Higher order
finiteness of the Higgs mass, where supersymmetry is broken in the
bulk by Scherk-Schwarz boundary conditions \cite{ss}, has been
discussed in \cite{quiros}.}. Towards the end of Section IV, we
briefly remark on the numerical implications of any possible
connection between $M_S$ and $R$.

\item The KK equivalent of Eq.~(\ref{epatop}), which captures the KK
contribution arising from the top quark chiral hypermultiplet, is then
given by 
\begin{equation}
\label{kkepatop}
\Delta V_t^n = {3 \over {32\pi^2}}
\left[m_{\tilde t_1^n}^4 \left(\ln{m_{\tilde
    t_1^n}^2\over{Q^2}}-{3\over2}\right)
      + m_{\tilde t_2^n}^4 \left(\ln{m_{\tilde
          t_2^n}^2\over{Q^2}}-{3\over2}\right)
      -2 m_{t^n}^4
\left(\ln{m_{t^n}^2\over{Q^2}}-{3\over2}\right)\right] , 
\end{equation}
where the field dependent KK masses are given by $m_{\tilde
t_{1}^n}^2 = m_{\tilde t_{1}}^2 + n^2/R^2$, $m_{\tilde t_{2}^n}^2
= m_{\tilde t_{2}}^2 + n^2/R^2$, and $m_{t^n}^2 = m_t^2 +
n^2/R^2$. The field dependence is hidden inside the zero mode
masses, as illustrated in Eqs.~(\ref{mthmbh}), (\ref{tsquark})
and (\ref{bsquark}). The corresponding contribution triggered by
the bottom quark hypermultiplet,  $\Delta V_b^n$, can be written
{\em mutatis mutandis}.

\item We now calculate the KK loop contribution to the neutral
scalar mass matrix. The procedure will be exactly the same as
that followed  for the 4d MSSM scenario in the previous section.
Since we are going to treat the radiatively corrected physical
$m_A$ as an input parameter, we concentrate only on the CP-even
mass matrix.  We first take another look at the expressions of
the different $\Delta_{ij}$, assembled in Eq.~(\ref{delta}),
calculated in the context of the 4d MSSM. The prefactors like
$m_t^4$ or $m_b^4$ originated by the action of double
differentiation on the field dependent squark or quark masses.
Recall that the squark and quark masses are (quadratically)
separated by the soft supersymmetry breaking mass-squares which
are {\em not} field dependent.  So, irrespective of whether we
double-differentiate the squark or quark masses we get either the
top or bottom quark Yukawa coupling\footnote{This also indicates
that by fixing the first and second generation matter superfields
at the brane we have not made any numerically serious compromise
as otherwise their contributions would have been adequately
suppressed on account of their small Yukawa couplings.}.  In the
same way, the KK mass-squares are separated from the zero mode
mass-squares by a field independent quantity $n^2/R^2$.
Therefore, the expressions for $(\Delta_{ij})^n$, the radiative
corrections from the $n$th KK level, continue to have the zero
mode quark masses $m_t^4$ or $m_b^4$ as prefactors, but now the
arguments of the other functions contain the corresponding KK
masses.
\end{enumerate} 

We are now all set to write down the expressions for different
$(\Delta_{ij})^n$ for $n \neq 0$. They are given by 
\begin{eqnarray}
\label{deltan}
(\Delta_{11}^t)^n &=& {m_t^4\over{{\rm sin}^2\beta}}\left(\mu (A_t+\mu
{\rm cot}\beta)\over{m_{\tilde t_1^n}^2 - m_{\tilde
t_2^n}^2}\right)^2g(m_{\tilde t_1^n}^2,m_{\tilde t_2^n}^2) ,\nonumber \\
(\Delta_{12}^t)^n &=& {m_t^4\over{\sin^2\beta}}{\mu (A_t+\mu 
\cot\beta)\over{m_{\tilde t_1^n}^2 - m_{\tilde t_2^n}^2}}\left[\ln
{{m_{\tilde t_1^n}^2}\over{m_{\tilde t_2^n}^2}}+{A_t(A_t+\mu \cot
\beta)\over{m_{\tilde t_1^n}^2 - m_{\tilde t_2^n}^2}}g(m_{\tilde
t_1^n}^2,m_{\tilde t_2^n}^2)\right] , \nonumber\\
(\Delta_{22}^t)^n& =& {m_t^4\over{{\rm sin}^2\beta}}\left[{
\ln}{{m_{\tilde t_1^n}^2}{m_{\tilde t_2^n}^2}\over{m_{t^n}^4}} +
{2A_t(A_t+\mu \cot\beta)\over{m_{\tilde t_1^n}^2 -m_{\tilde
t_2^n}^2}} \ln {{m_{\tilde t_1^n}^2}\over{m_{\tilde
t_2^n}^2}} + \left(A_t(A_t+\mu
\cot\beta)\over{m_{\tilde t_1^n}^2 - m_{\tilde
t_2^n}^2}\right)^2g(m_{\tilde t_1^n}^2,m_{\tilde t_2^n}^2) \right] ,
\nonumber\\
(\Delta_{11}^b)^n& =& {m_b^4\over{{\rm cos}^2\beta}}\left[
\ln{{m_{\tilde b_1^n}^2}{m_{\tilde b_2^n}^2}\over{m_{b^n}^4}} +
{2A_b(A_b+\mu \tan\beta)\over{m_{\tilde b_1^n}^2 -m_{\tilde
b_2^n}^2}} \ln{{m_{\tilde b_1^n}^2}\over{m_{\tilde
b_2^n}^2}} + \left(A_b(A_b+\mu
\tan\beta)\over{m_{\tilde b_1^n}^2 - m_{\tilde
b_2^n}^2}\right)^2g(m_{\tilde b_1^n}^2,m_{\tilde b_2^n}^2)\right] ,
\nonumber \\
(\Delta_{12}^b)^n &=& {m_b^4\over{\cos^2\beta}}{\mu (A_b+\mu 
\tan\beta)\over{m_{\tilde b_1^n}^2 - m_{\tilde b_2^n}^2}}\left[
\ln{{m_{\tilde b_1^n}^2}\over{m_{\tilde b_2^n}^2}}+{A_b(A_b+\mu 
\tan\beta)\over{m_{\tilde b_1^n}^2 - m_{\tilde b_2^n}^2}}g(m_{\tilde
b_1^n}^2,m_{\tilde b_2^n}^2)\right] ,\\
(\Delta_{22}^b)^n &=& {m_b^4\over{\cos^2\beta}}\left(\mu (A_b+\mu
\tan\beta)\over{m_{\tilde b_1^n}^2 - m_{\tilde
b_2^n}^2}\right)^2g(m_{\tilde b_1^n}^2,m_{\tilde b_2^n}^2) .\nonumber 
\end{eqnarray} 
Now we have to add the $(\Delta^t)^n$ and $(\Delta^b)^n$ matrices to
the one-loop corrected (from zero modes only) mass matrix in
Eq.~(\ref{msqeven}), sum over $n$, and then diagonalise to obtain the
eigenvalues $m_h^2$ and $m_H^2$. The KK radiative corrections decouple
in powers of $(R^2/n^2)$. To provide intuition to the expressions in
Eq.~(\ref{deltan}), we display below the approximate formulae for
$(\Delta^t)^n$ in leading powers of $(R^2/n^2)$:
\begin{eqnarray}
\label{apdeltan}
(\Delta_{11}^t)^n & = & - \frac{1}{6}
\left(\frac{R^4}{n^4}\right)
\frac{m_t^4}{\sin^2\beta} \left[ \mu(A_t+\mu\cot\beta) \right]^2 ,
\nonumber \\ (\Delta_{12}^t)^n & = & \left(\frac{R^2}{n^2}\right)
\frac{m_t^4}{\sin^2\beta} \mu(A_t+\mu\cot\beta) , \\
(\Delta_{22}^t)^n & = & \left(\frac{R^2}{n^2}\right)
\frac{m_t^4}{\sin^2\beta} \left[(m_{\tilde t_1}^2 + m_{\tilde t_2}^2 -
2m_t^2) + 2 A_t(A_t+\mu\cot\beta)\right] . \nonumber 
\end{eqnarray} 
Similar expressions for $(\Delta^b)^n$ can be written, with
appropriate replacements like $m_t \leftrightarrow m_b$, $\cot\beta 
\leftrightarrow \tan\beta$, etc. So, in the absence of any left-right
scalar mixing, the KK contribution to $\Delta m_h^2$ is controlled by
$R^2 (m^2_{\tilde t} - m_t^2)/n^2$ and its higher powers.

{\bf Six dimensional scenario:}~ For the 6d scenario we follow the
compactification on a chiral square, as done in \cite{sixd}, which
admits zero mode chiral fermions. The two extra spatial coordinates
$(y_1,y_2)$ are compactified on a square of side length $L$, such that
$0 < y^1,~y^2 < \pi R (\equiv L)$. The boundary condition is the
identification of the two pairs of adjacent sides of the squares such
that the values of a field at two identified points differ by a phase
($\theta$). Nontrivial solutions exist when $\theta$ takes four
discrete values $(n\pi/2)$ for $n=0,1,2,3$ and the zero modes appear
when $n=0$. What matters to our calculation in this paper is the
structure of the KK masses, a generic pattern of which is given by
\begin{equation}
\label{kksixd} 
m_{j,k}^2 = m_0^2 + \frac{j^2+k^2}{R^2} ,  
\end{equation} 
where $j,k$ are integers such that $j \geq 0$ and $k \geq 0$.  We
display in Table \ref{sixdm} the KK mass spectrum (neglecting the zero
mode mass $m_0$ for simplicity of presentation while in the actual
calculation we do keep it). The formalism we developed for 5d will
simply go through for 6d. More concretely, the structure of
Eqs.~(\ref{kkepatop}) and (\ref{deltan}) would remain the same in 6d,
only that one should now read $n \Rightarrow (j,k)$.  The numerical
impact in the two cases obviously differ, as we shall witness in the
next section\footnote{Admittedly, the 6d sum is logarithmically
sensitive to the cutoff. The low-lying KK states we include reflect
the dominant contribution to the Higgs mass shift. We thank Anindya
Datta for raising the 6d divergence issue.}.
%%%%%%%%%%%%%%%%%%%%%%%%%%%%%%%%%%%%%%%%%%%%%%%%%%%%%%%%%%
\begin{center}
\begin{table}
\begin{tabular}{|l|c|c|c|c|c|c|c|c|c|} \hline
$(j,k)$&$1,0$&(1,1)&(2,0)&(2,1) or (1,2)&(2,2)&(3,0)&(3,1) or
(1,3)&(3,2) or (2,3)&(4,0)\\ \hline $m_{j,k}$&1&$\sqrt 2$&2&$\sqrt
5$&$2\sqrt 2$&3&$\sqrt {10}$& $\sqrt {13}$&4 \\\hline
\end{tabular}
\caption{{\em 6d scenario mass spectrum in $(1/R)$
units, neglecting the zero mode mass}.}
\label{sixdm}
\end{table}
\end{center}
%%%%%%%%%%%%%%%%%%%%%%%%%%%%%%%%%%%%%%%%%%%%%%%%%%%%%

\section{Results}
In this section we explore the consequences of the
extra-dimensional contributions to the Higgs mass encoded in the
exact one-loop expressions in Eq.~(\ref{deltan}). But to start
with, to get a feel for the numerical impact of the extra
dimensions, consider the scenario pared down to its bare minimum
by assuming that left-right scalar mixing ingredients are
vanishing, i.e., $\mu = A_t = A_b = 0$. This leads to two
degenerate stop squarks: $m_{\tilde{t}}^2 = M_S^2 + m_t^2$. Then,
for a moderate $\tan\beta$, 
\begin{equation} 
\label{approxmh}
\Delta m_h^2 ~(n=0) \sim \frac{3 m_t^4 }
{2\pi^2 v^2} \ln\left(1 + \frac{M_S^2}{m_t^2}\right); ~~~
\Delta m_h^2 ~(n \neq 0) \sim \frac{3 m_t^4 }{2\pi^2 v^2} 
\frac{(M_S R)^2}{n^2}   .
\end{equation}
Indeed, non-zero trilinear and $\mu$ terms would complicate the
expressions, yet Eq.~(\ref{approxmh}) provides a good intuitive
feel for our results displayed through the different plots. The
expected decoupling of extra-dimensional effects in the $1/R
\rightarrow \infty$ limit is transparent in Eq. (\ref{approxmh}),
leaving the logarithmic dependence on the supersymmetry scale,
$M_S$.

As stressed already, the primary emphasis in this work is to
examine the effect of extra dimensions on the upper bound of
$m_h$. In 4d supersymmetry it is usual to choose the pseudoscalar Higgs
mass, $m_A$, as a free parameter and exhibit $m_h$ as its
function. This has been done for the extra-dimensional MSSM
models in Figs. 1 (5d case) and 2 (6d case). Let us discuss them
in turn.

In these and the subsequent figures, the parameters involved are
chosen as follows:\\
(a)~ $m_Q = m_U = m_D \equiv M_S$, which is a common soft
supersymmetry breaking mass. Several values of 
$M_S$ have been chosen in the figures to depict its impact. \\
(b)~ The trilinear scalar couplings  $A_t$ and $A_b$ are varied in
the range $[0.8 - 1.2]~M_S$. This results in bands in the
figures. We have found that the results are not particularly
sensitive to $\mu$ and we hold it fixed at 200 GeV. Also, sign
flips in the trilinear couplings do not change the results.\\
(c)~ The stop and sbottom (zero mode) mass eigenvalues are
calculated from the diagonalisation of matrices in
Eqs.~(\ref{tsquark}) and (\ref{bsquark}) after setting the Higgs
fields to their VEVs. For a chosen value of $\tan\beta$ and $M_S$,
those eigenvalues will vary in a range in accord with the variation
of $A_t$ and $A_b$ stated above. \\ 
(d)~ Since we are interested in probing the upper limit of the
lightest Higgs, we maximize its {\em tree} level mass as much as
possible.  For displaying our results we have fixed $\tan\beta =
10$, a moderate value for which the tree level $m_h$ is almost close 
to $M_Z$.

In Fig.~1 we have displayed the result in the $m_h$-$m_A$ plane
for only one extra dimension. The dependence of $m_h$ on $m_A$ in
the MSSM case is mimicked in the extra-dimensional case and $m_h$
settles at its upper limit for $m_A$ greater than about 150 GeV.
In the left panel, $M_S$ has been fixed at 500 GeV.  As
anticipated, larger the value of $1/R$ smaller is the extra
dimensional impact. The 4d MSSM case corresponds to $1/R \to
\infty$. The width of each band reflects the variation of the
trilinear parameters in the zone mentioned above.  For the chosen
supersymmetry parameters, the maximum value of $m_h$ is a little
below 125 GeV for the 4d MSSM case while for the extra-dimensional
situation it is enhanced to above 135 (130) GeV for $1/R$ = 600
GeV (1 TeV). In the right panel, the dependence on $M_S$ is
exhibited holding $1/R$ at 1 TeV.  Clearly, a larger 
$M_S$ results in  bigger radiative corrections -- recall Eq.
(\ref{approxmh}) -- both from the zero mode as well as from the
KK modes.

Fig.~2 is a 6d version of  Fig.~1. While the pure 4d MSSM band
remains the same, the KK radiative effects are larger now due to
the denser KK spectrum in the 6d case, specified by two sets of
integers $j$ and $k$, as shown in Table ~1.  Quantitatively, for
an $1/R$ of 600 GeV (1 TeV), $m_h$ can now be as heavy as 195
(155) GeV, to be compared with 125 GeV in 4d MSSM for these
parameter values.

As mentioned earlier, the current lower bound on $m_h$ of 114.5
GeV excludes low values of $\tan \beta$ in the 4d MSSM. It is
expected that in the extra-dimensional scenarios some of this
excluded range of $\tan \beta$ will make it into the allowed
zone. In Fig.~3, we have shown the variation of $m_h$ with
respect to $\tan\beta$ (for low values) to illustrate this
effect. For the 5d case (left panel), $1/R$ of even 1.2 TeV
eases the tension somewhat while for $1/R$ of 600 GeV the effect
is very prominent.  For 6d (right panel), the extra-dimensional
contributions are further enhanced and the restriction on $\tan
\beta$ is essentially entirely lifted.  We should recall that
$\tan\beta$ enters in the Higgs couplings to other particles and
so the above result has significant bearing on collider searches
of supersymmetry.

So far, we have exhibited results for a few choices of the
compactification scale, $1/R$. Fig.~4 demonstrates how the
KK-induced radiative correction depends on $1/R$ for the 5d (left
panel)  and  6d (right panel) scenarios. If the Higgs boson is
detected at the LHC then using these figures one can gain a
handle on $1/R$ dependent on the supersymmetry parameters like
$M_S$.  The decoupling behaviour as $1/R$ increases is in
agreement with expectation.

We have also studied, in passing, the possibility that the soft
supersymmetry breaking scale arises from compactification
(e.g. through the Scherk-Schwarz mechanism \cite{ss}). Let us suppose
$M_S = C/R$, where $C$ is an order one dimensionless constant. Since
we are interested in weak scale supersymmetry breaking, we keep $1/R$
around a few hundred GeV to a TeV. In this region, the radiative
correction roughly depends on $M_S$ and $R$ only through their product
($\equiv C$), and for a choice of $C \in [0.5-2.0]$, the upper limit
on the lightest Higgs mass turns out to be in the range $m_h \in
(150-230)$ GeV (5d) and $(200-450)$ GeV (6d).

It may bear mentioning again that in these calculations we have
retained the loop conntributions from the $t$ and $b$ quarks only. The
other quarks and gauge bosons make negligible impact.  Also, we have
dealt only with real MSSM parameters and limited our studies up to
one-loop KK contributions. We have not, therefore, included either the
two-loop improvements of the 4d MSSM calculations or the numerical
effects of the phases associated with complex MSSM parameters in our
discussions (for a recent survey, see \cite{Heinemeyer:2007aq}).

\section{Conclusions}
One of the virtues for which supersymmetry stands out as a leading
candidate of physics beyond the SM is that it sets an upper bound
on the Higgs mass.
The lightest neutral Higgs mass, $m_h$, could at most be $M_Z$ at the
tree level, but is pushed further obeying a definite relation,
obtained from quantum corrections, involving $m_h$, $m_t$ and the stop
squark mass, $m_{\tilde{t}}$. The sensitivity of this correction to
$m_{\tilde{t}}$ is only logarithmic. Consequently, a firm prediction
results, namely, that $m_h ~\ltap ~135$ GeV in MSSM for
$m_{\tilde{t}}~\ltap ~{\cal{O}}$(1 TeV). This is regarded as a
critical test of supersymmetry and is naturally high on the
agenda of the upcoming LHC experiments. In this paper, we have probed
how much this upper limit could be relaxed, should the MSSM be embedded
in one ($S^1/Z_2$) or two ($T^2/Z_4$) extra dimensions. We highlight
our main findings:
\begin{enumerate}
\item The KK towers of the  top quark and stop squarks provide a
positive contribution to $m_h^2$ raising it by several tens of GeV.
If we ignore left-right scalar mixing and assume moderate
$\tan\beta \sim (5-10)$, then using Eq.~(\ref{approxmh}) and
summing over all the KK modes, we obtain $\Delta m_h^2 ({\rm KK})
\sim (60~{\rm GeV})^2 \times (M_S R)^2$. This is a 5d result.
Including the left-right scalar mixings, i.e., non-zero $\mu$ and
trilinear parameters, somewhat enhances the magnitude of the
correction (see Fig.~1). As in the case of 4d MSSM, here too the size
of the correction is controlled by the large top Yukawa coupling.

\item If we consider a 6d theory with two extra dimensions
compatcified on a chiral square, whose motivations have been
mentioned earlier, the correction gets sizably enhanced (see
Fig.~2), compared to 5d, due to a denser packing of KK states,
which are now fixed by two independent KK numbers.

\item Non-observation of a Higgs boson weighing below 114.5 GeV
disfavours low $\tan\beta$ in 4d MSSM. Some part of this 
region can be revived by extra dimensional embedding (see
Fig.~3).

\item The 4d MSSM relationship between the lightest neutral Higgs mass
and the stop squark mass is extremely profound in the sense that its
specific form does not depend on the supersymmetry breaking
mechanism. If supersymmetry is embedded in extra dimension(s) and,
with some cooperation from Nature, the KK states happen to be light
enough to mark their imprints on the LHC data recorder, then the
relationship between the stop mass and the Higgs mass alters in a
numerically significant way (see Fig.~4).  

\end{enumerate}

\noindent{ \bf Acknowledgements:}
G.B. acknowledges hospitality at the Abdus Salam ICTP, Trieste, during
a part of the work, and thanks G. Senjanovi\'c for his critical
remarks.  G.B. also recognizes the contribution of the ever refreshing
`Osmizzas' towards this paper and thanks Angelo for accompanying him
there during his stay at Trieste.

\newpage
%%%%%%%%%%%%%%%%%%%%%%%%%%%%%%%%%%%%%%%%%%%%%%%%%%%%%%%%%%%%%%%%%%%
\begin{figure}[t]
  \centering
  \includegraphics[width=0.9\textwidth,height=0.32\textheight]
  {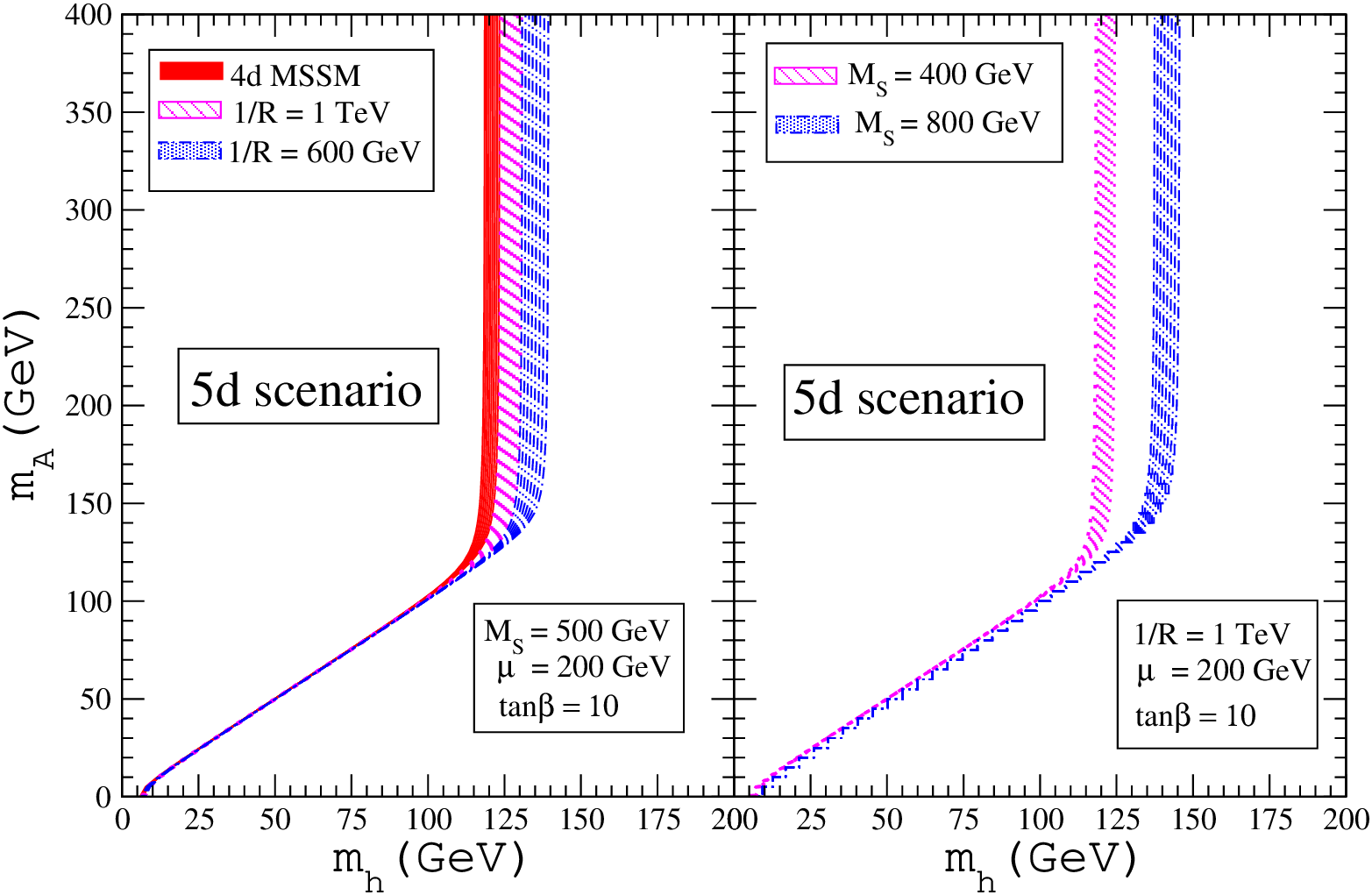} \caption{\em The variation of $m_h$ with $m_A$ in the
  5d MSSM for different choices of the supersymmetry breaking scale
  ($M_S$) and the compactification radius ($R$). The width of each
  band corresponds to the variation of $A_t$ and $A_b$ in the range
  $(0.8 - 1.2)M_S$ (see text).}
%  \label{}
\end{figure}
%%%%%%%%%%%%%%%%%%%%%%%%%%%%%%%%%%%%%%%%%%%%%%%%%%%%%%%%%%%%%%%%%%%

%%%%%%%%%%%%%%%%%%%%%%%%%%%%%%%%%%%%%%%%%%%%%%%%%%%%%%%%%%%%%%%%%%%
\begin{figure}[b]
  \centering
  \includegraphics[width=0.9\textwidth,height=0.32\textheight]
  {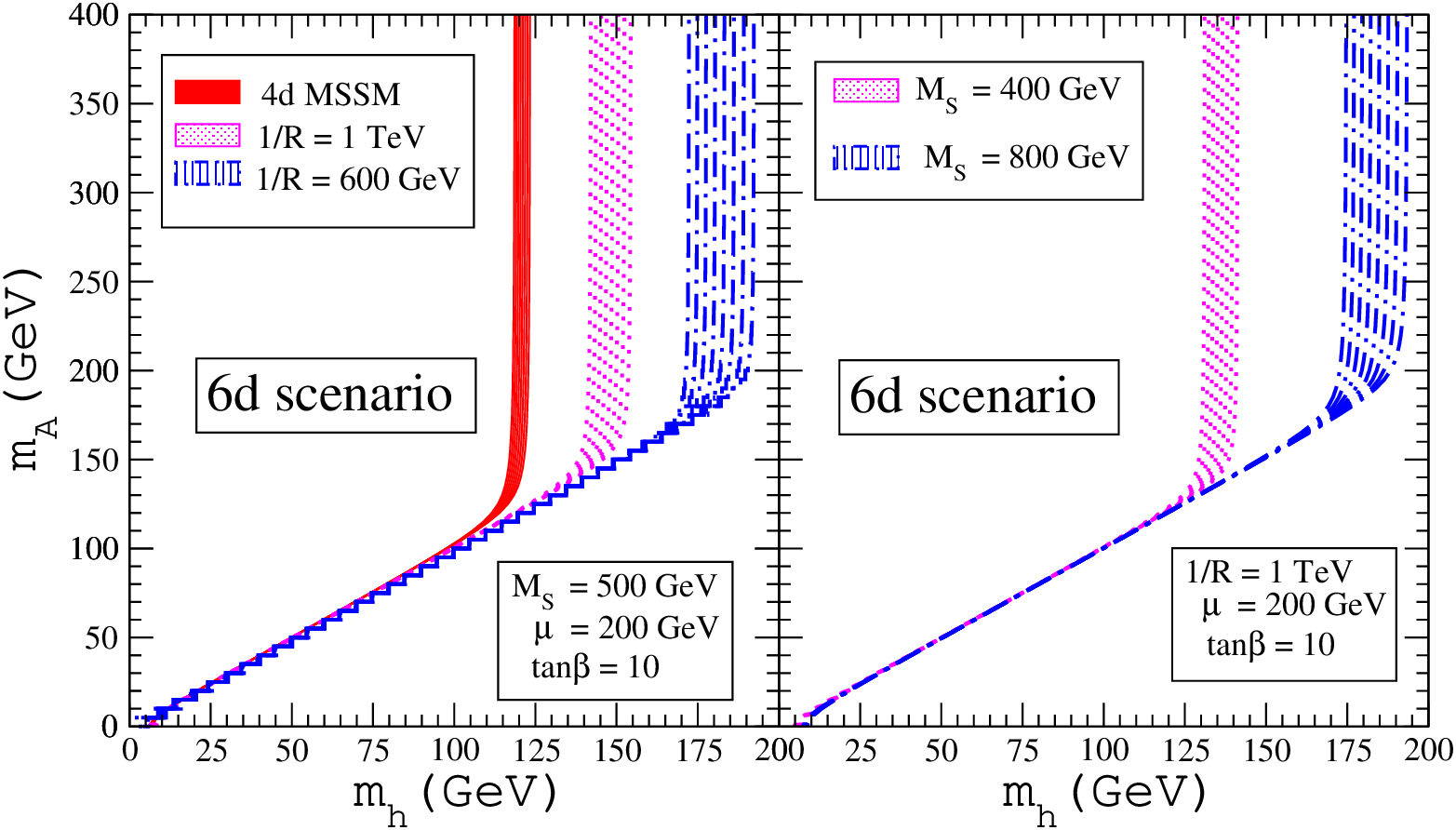}
   \caption{\em As in Fig.~1 but for 6d MSSM.}
%  \label{}
\end{figure}
%%%%%%%%%%%%%%%%%%%%%%%%%%%%%%%%%%%%%%%%%%%%%%%%%%%%%%%%%%%%%%%%%%%

%\newpage

%%%%%%%%%%%%%%%%%%%%%%%%%%%%%%%%%%%%%%%%%%%%%%%%%%%%%%%%%%%%%%%%%%%
\begin{figure}[t]
  \centering
  \includegraphics[width=0.9\textwidth,height=0.30\textheight]
  {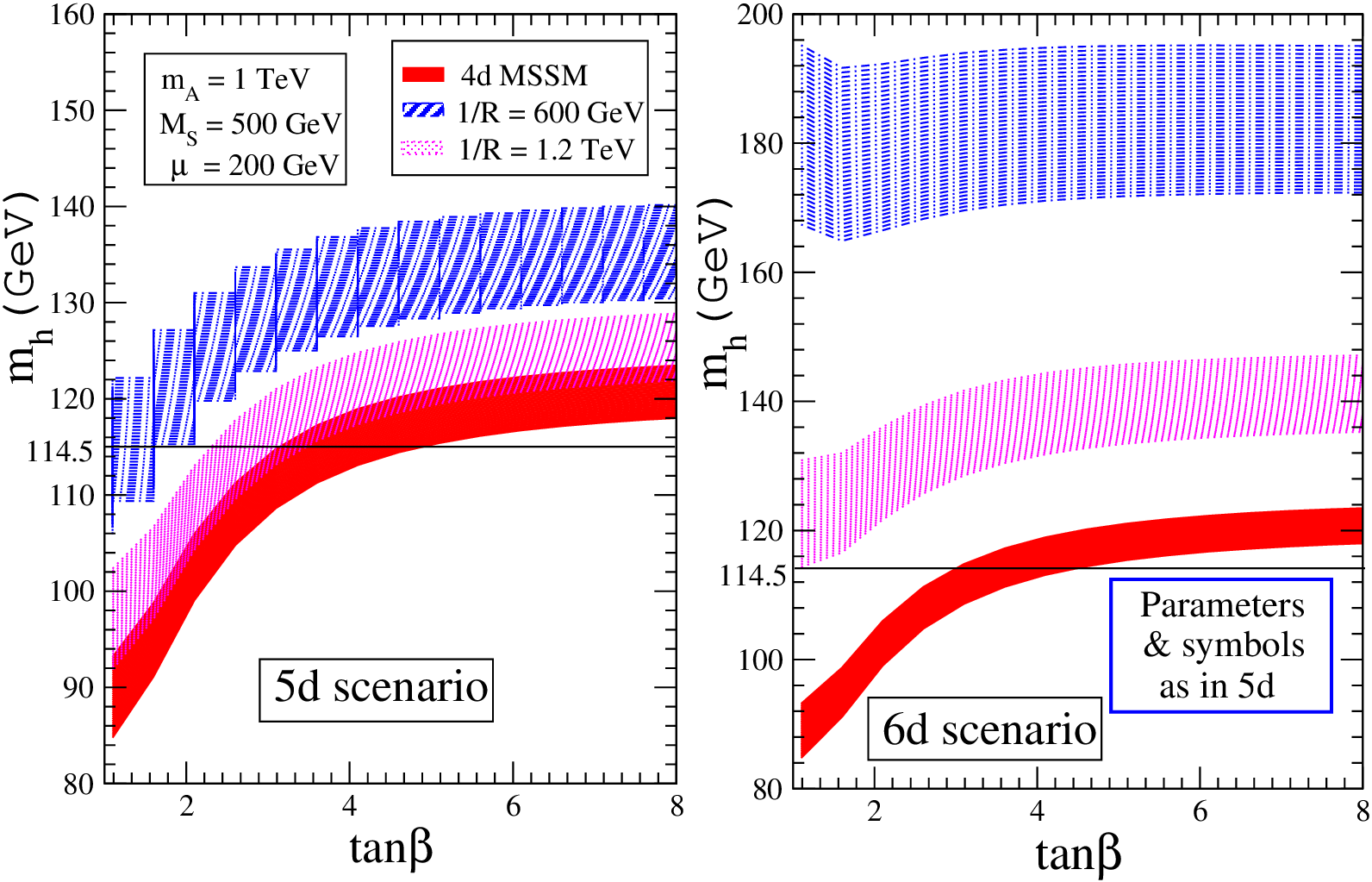} \caption{\em The dependence of $m_h$ on $\tan\beta$
  (zoomed for the low values) in 5d (left panel) and 6d (right panel)
  MSSM.  The width of each band corresponds to the variation of $A_t$
  and $A_b$ in the range $(0.8 - 1.2)M_S$.}
%  \label{}
\end{figure}
%%%%%%%%%%%%%%%%%%%%%%%%%%%%%%%%%%%%%%%%%%%%%%%%%%%%%%%%%%%%%%%%%%%

%%%%%%%%%%%%%%%%%%%%%%%%%%%%%%%%%%%%%%%%%%%%%%%%%%%%%%%%%%%%%%%%%%%
\begin{figure}[b]
  \centering
  \includegraphics[width=0.9\textwidth,height=0.30\textheight]
  {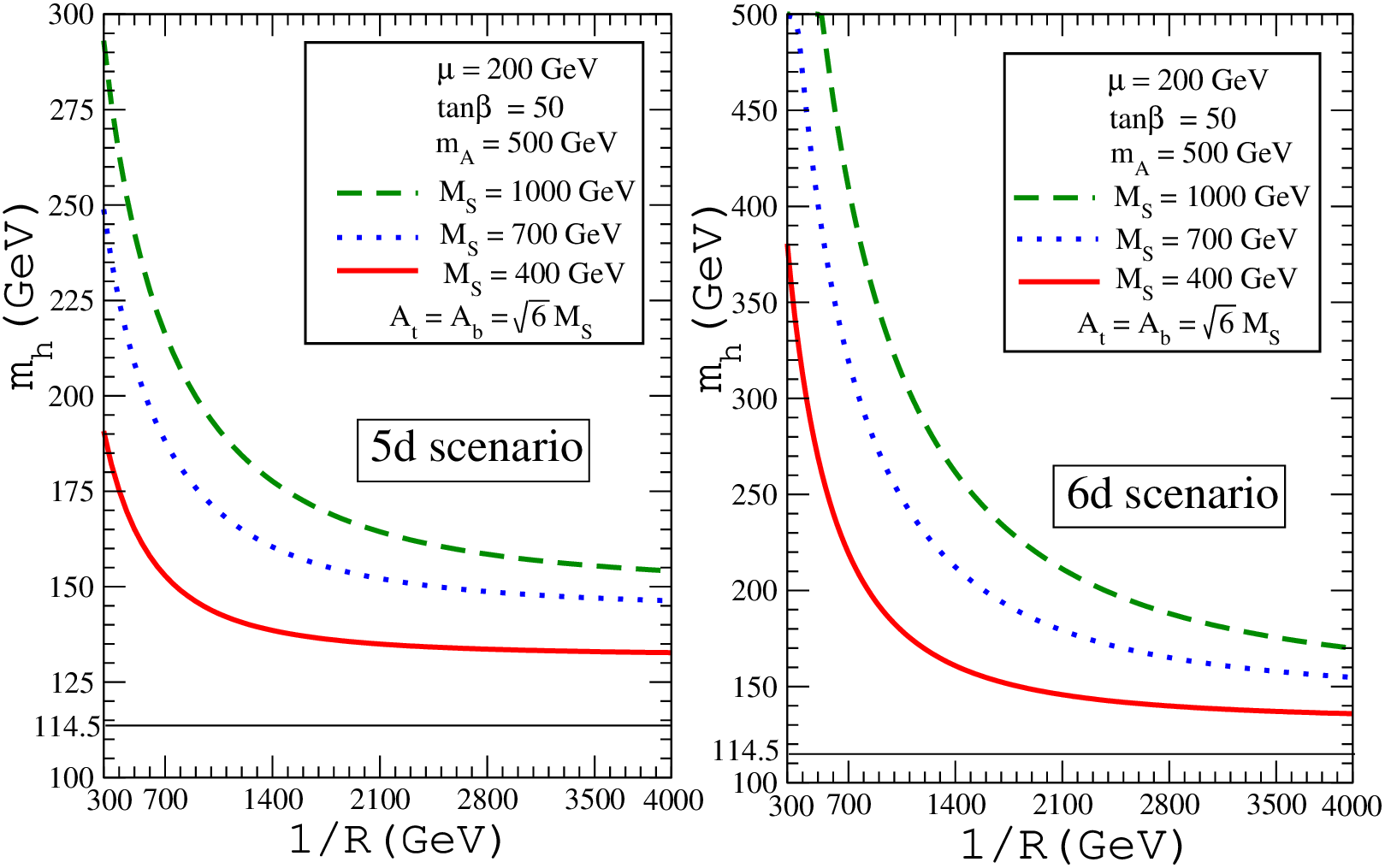} \caption{\em The dependence of $m_h$ on $1/R$
  for different choices of $M_S$ for 5d (left panel) and 6d (right
  panel) cases. The ratio $\sqrt{6}$ between $A_t (= A_b)$ and $M_S$
  maximises the trilinear contribution (see, Drees, Godbole, Roy in
  \cite{books}).} 
%  \label{}
\end{figure}
%%%%%%%%%%%%%%%%%%%%%%%%%%%%%%%%%%%%%%%%%%%%%%%%%%%%%%%%%%%%%%%%%%%

\end{document}